\documentclass[11pt,twoside]{article}
\usepackage{asp2014}

\aspSuppressVolSlug
\resetcounters

\providecommand{\tightlist}{%
  \setlength{\itemsep}{0pt}\setlength{\parskip}{0pt}}

\begin{document}

\title{Towards new solutions for scientific computing: the case of Julia}

\author{Maurizio~Tomasi,$^1$ and Mos\'e~Giordano$^2$
  \affil{$^1$Universit\`a{} degli Studi, Milano, Italy; \email{maurizio.tomasi@unimi.it}}
  \affil{$^2$Universit\`a{} del Salento, Lecce, Italy; \email{mose.giordano@le.infn.it}}}

\paperauthor{Maurizio~Tomasi}{maurizio.tomasi@unimi.it}{0000-0002-1448-6131}{Universit\`a{} degli Studi di Milano}{Dipartimento di Fisica ``A.~Pontremoli''}{Milan}{}{20133}{Italy}
\paperauthor{Mos\'e~Giordano}{mose.giordano@le.infn.it}{0000-0002-7218-2873}{Universit\`a{} del Salento}{Dipartimento di Matematica e Fisica ``E.~De Giorgi''}{Lecce}{}{73100}{Italy}


\begin{abstract}This year marks the consolidation of Julia (https://julialang.org/), a
programming language designed for scientific computing, as the first
stable version (1.0) has been released, in August 2018. Among its main
features, expressiveness and high execution speeds are the most
prominent: the performance of Julia code is similar to statically
compiled languages, yet Julia provides a nice interactive shell and
fully supports Jupyter; moreover, it can transparently call external
codes written in C, Fortran, and even Python and R without the need of
wrappers. The usage of Julia in the astronomical community is growing,
and a GitHub organization named JuliaAstro takes care of coordinating
the development of packages. In this paper we present the features and
shortcomings of this language, and discuss its application in astronomy
and astrophysics.\end{abstract}

\hypertarget{introduction}{%
\section{Introduction}\label{introduction}}

Julia \citep{Bezanson2017} is a programming language that has recently
reached its first stable milestone: version 1.0 has been released in
August 2018, and the language specification has been freezed. Julia
provides a number of features that makes it extremely interesting for
astrophysics, astronomy, and scientific applications in general:

\begin{itemize}
\tightlist
\item
  Nice and simple syntax, similar to Matlab's.
\item
  The speed of Julia codes often matches the speed of other languages
  used for High Performance Computing (HPC), namely C, C++, and Fortran,
  thanks to a number of features: type inference, Just-In-Time
  compilation, use of LLVM to produce optimized machine code;
\item
  Native support for vectors, matrices, and tensors;
\item
  Support for missing values (using the keyword \texttt{missing}),
  useful when dealing with data acquired using real-world experiments;
\item
  First-class support for many numeric types, apart from integers and
  floating-point numbers: rationals, complex numbers,
  arbitrary-precision numbers.
\item
  Symbolic computation (e.g., estimation of analytical derivatives) is
  easy to implement;
\item
  Easy to call functions defined in dynamic libraries, using the
  \texttt{ccall} function;
\item
  Ability to import packages written in Python or R; several wrappers to
  well-known Python libraries are available (e.g., \texttt{PyPlot.jl}
  wraps Matplotlib).
\end{itemize}

\hypertarget{features-of-julia}{%
\section{Features of Julia}\label{features-of-julia}}

\hypertarget{compilation-model}{%
\subsection{Compilation model}\label{compilation-model}}

Julia compiles functions the first time they are executed. The
compilation depends on the type of the function parameters, as shown in
this example:

\begin{verbatim}
f(x) = 2x + 1   # Define a function
f(1)            # Compile f assuming an integer argument
f(1.0)          # Compile again f assuming a float argument
f(3)            # No compilation is necessary, as 3 is an int
\end{verbatim}

\hypertarget{operations-on-arrays-matrices-and-tensors}{%
\subsection{Operations on arrays, matrices, and
tensors}\label{operations-on-arrays-matrices-and-tensors}}

Julia's arrays are similar to Fortran's:

\begin{enumerate}
\def\labelenumi{\arabic{enumi}.}
\tightlist
\item
  Indices start from 1;
\item
  Arrays are stored in column-major order;
\item
  The compiler is able to propagate operators and functions to arrays,
  performing loop fusion.
\end{enumerate}

The latter point is particularly important. If \texttt{a}, \texttt{b},
\texttt{c}, and \texttt{result} are arrays of the same size, the
statement \texttt{result\ =\ a\ +\ b\ +\ c} in Fortran corresponds to
one \texttt{do} loop. On the other side, the same code in Python applied
on NumPy arrays is equivalent to the application of \emph{three}
\texttt{for}-loop cycles, because NumPy is not able to perform\footnote{This
  limitation can be circumvented by other libraries, like WeldNumPy
  (\href{https://www.weld.rs/weldnumpy/}{www.weld.rs/weldnumpy}), Numba
  (\href{https://numba.pydata.org/}{numba.pydata.org}), or Cython
  (\href{https://cython.org/}{cython.org}).} \emph{loop fusion}, i.e.,
the combination of several \texttt{for} loops into one.

Loop fusion is an important feature for HPC languages. Julia provides
loop fusion through the so-called \emph{dotted operators}: if
\texttt{\#} is a two-argument operator, \texttt{.\#} applies the
operator to all the elements of the two arrays. Therefore, in Julia the
code \texttt{result\ .=\ a\ .+\ b\ .+\ c} is equivalent to the Fortran
code \texttt{result\ =\ a\ +\ b\ +\ c}. Julia's approach is more
general, as this applies to custom operators and functions as well:

\begin{verbatim}
++(a::Real, b::Real) = 2a + b      # Custom operator
3 ++ 4                             # Result: 10
[3, 4] .++ [4, 7]                  # Result: [10, 15]
f(x::Real) = 3x^2                  # Custom function
f.([3, 6, 5])                      # Result: [27, 108, 75]
\end{verbatim}

\hypertarget{homoiconicity}{%
\subsection{Homoiconicity}\label{homoiconicity}}

Julia provides the syntax for manipulating its own code with the same
syntax used to manipulate variables. This feature, called
\emph{homoiconicity} (``same representation''), is inspired by LISP-like
languages, and it has several applications in the domain of symbolic
analysis (e.g., automatic computation of analytical derivatives). An
interesting applications of homoiconicity in Julia is provided by the
Zygote package \citep{Innes2018}, which is able to perform automatic
symbolic differentiation at compile time:

\begin{verbatim}
julia> using Zygote
julia> f(x) = 2x + 1
julia> @code_llvm f'(0)
; Function #68
; Location: /somewhere/interface.jl:49
define i64 @"julia_#68_37159"(i64) {
top:
  ret i64 2 # Return 2 immediately (the derivative is a constant)
}
\end{verbatim}

\hypertarget{julia-in-astronomy}{%
\section{Julia in Astronomy}\label{julia-in-astronomy}}

\hypertarget{juliaastro}{%
\subsection{JuliaAstro}\label{juliaastro}}

The JuliaAstro GitHub organization
(\href{https://github.com/JuliaAstro}{github.com/JuliaAstro}) collects
all the packages related to astronomy developed for Julia. At the time
of writing (November 2018), the packages are the following:

\begin{itemize}
\tightlist
\item
  \texttt{AstroImages.jl}: Visualization of astronomical images;
\item
  \texttt{AstroLib.jl}: Bundle of small astronomical and astrophysical
  routines;
\item
  \texttt{AstroTime.jl}: Astronomical time keeping;
\item
  \texttt{Cosmology.jl}: Library of cosmological functions;
\item
  \texttt{DustExtinction.jl}: Models for the interstellar extinction due
  to dust;
\item
  \texttt{ERFA.jl}: Wrapper to \texttt{liberfa}\footnote{\href{https://github.com/liberfa/erfa}{github.com/liberfa/erfa}.
    This is a BSD-licensed replica of the SOFA library
    (\href{http://www.iausofa.org/}{www.iausofa.org}).};
\item
  \texttt{EarthOrientation.jl}: Earth orientation parameters from IERS
  tables;
\item
  \texttt{FITSIO.jl}: Flexible Image Transport System (FITS) file
  support;
\item
  \texttt{LombScargle.jl}: Compute Lomb-Scargle periodogram;
\item
  \texttt{SPICE.jl}: Julia wrapper for NASA NAIF's SPICE toolkit;
\item
  \texttt{SkyCoords.jl}: Support for astronomical coordinate systems;
\item
  \texttt{UnitfulAstro.jl}: An extension of \texttt{Unitful.jl} (a
  package to attach measure units to variables) for astronomers;
\item
  \texttt{WCS.jl}: Astronomical World Coordinate Systems library.
\end{itemize}

\hypertarget{simulating-a-cmb-space-mission}{%
\subsection{Simulating a CMB space
mission}\label{simulating-a-cmb-space-mission}}

One of us (MT) has had the opportunity to use Julia in a few studies
involving the design of a CMB space mission (CORE, PICO, and LiteBIRD).
These studies involved the simulation of the operations needed to
observe the sky, and they required the generation of simulated noisy
data timelines acquired by instruments mounted onboard the spacecraft.
The quantity of data was of the order of hundreds of GB, and the
exploratory nature of the study made existing codes (developed in C++
for the Planck experiment) cumbersome to use, as they were conceived as
large monolithic programs meant to be ran end-to-end. A rewrite of some
modules in Julia provided similar performance (within 10\%) with the
existing C++ codes; moreover, the Julia codes were runnable in Jupyter
notebooks, thus allowing to interactively explore the parameter space
and ease data analysis.

\hypertarget{conclusions}{%
\section{Conclusions}\label{conclusions}}

Julia has several features that make it an interesting solution for
astronomical and astrophysical projects. It can achieve performance
similar to compiled languages, like C and Fortran, but it is
considerably more expressive and easy to use.

Nothwistanding the long list of interesting features, we believe it
would not be fair to omit some of Julia's most important shortcomings:

\begin{itemize}
\tightlist
\item
  Compilation times can be significant. Since compilation happens at
  runtime, a Julia script that calls several short functions can be
  noticeably slower than a similar script written in other compiled or
  interpreted languages.
\item
  The language is new, and there are not as many libraries as for other
  languages. Python, R, C, and Fortran library are easy to import;
  however, if a code heavily relies only on a few libraries, it is
  usually easier to just use the language for which these libraries were
  developed than wrapping everything in Julia.
\item
  It is still not possible to produce stand-alone executables. This
  makes code deployment more difficult.
\item
  As any new language, it is necessary to grasp a number of concepts
  before being fully productive with it. For instance, a programmer
  experienced in NumPy might find surprising that explicit \texttt{for}
  loop can be more performant than expressions involving broadcasting.
  (The repository
  \href{https://github.com/ziotom78/python-julia-c-}{github.com/ziotom78/python-julia-c-}
  provides an example.)
\end{itemize}

In the opinion of the authors, there are two contexts in astrophysical
data analysis where Julia can provide a significant advantage over
existing solutions:

\begin{itemize}
\tightlist
\item
  Analysis of large amounts of data, where no existing codes are
  available and the amount of calculations is significant. In this case,
  Julia codes can be as performant as other codes written using multiple
  libraries and languages: the typical case uses Python for most of the
  code and some optimized library (Numba, Fortran codes wrapped using
  \texttt{f2py}) for the most performance-critical routines. As an
  application of this use case we mention the Celeste project, which was
  able to load and process 178 TB of data from the SDSS catalogue in
  14.6 minutes across 8192 nodes \citep{2018arXiv180110277R}.
\item
  Existing codes are monolithic and difficult to use interactively, and
  the expense of rewriting code in Julia can be rewarded by the
  possibility to run the code interactively, either in Julia's command
  line or in Jupyter notebooks.
\end{itemize}

\acknowledgements We thank the Julia community at
\href{https://discourse.julialang.org/}{discourse.julialang.org} for
many useful discussions

\bibliography{O4-5}  

\end{document}